\documentclass{jetpl}
\twocolumn

\lat

\title{Pore formation phase diagrams for lipid membranes}

\rtitle{Pore formation phase diagrams\ldots}

\sodtitle{Pore formation phase diagrams for lipid membranes}

\author{S.\,I. Mukhin\thanks{e-mail: i.m.sergei.m@gmail.com}, B.\,B. Kheyfets}

\rauthor{S.\,I. Mukhin, B.\,B. Kheyfets}

\sodauthor{Mukhin, Kheyfets}

\address{Theoretical Physics and Quantum Technology Department, NUST ``MISIS'', 119049  Moscow, Russia}

\dates{ }{*}

\abstract{
  Critical lateral pressure for a pore formation and phase diagram of porous membrane are derived analytically as functions of the microscopic parameters of the lipid chains.  The derivation exploits path-integral calculation of the free energy of the ensembles  of semi-flexible strings and rigid rods that mimic the hydrophobic tails of lipids in the lipid bilayers and bolalipid membranes respectively. Analytical expressions for the area stretch/compressibility moduli of the membranes are derived in both models.
}


\begin{document}

\maketitle
\begin{figure}[tbp]
\includegraphics[width=\linewidth]{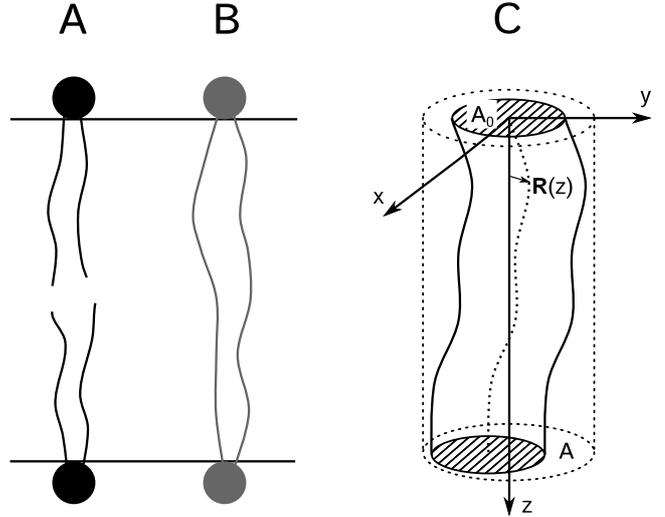}
\caption{Fig. 1. A: Monopolar lipid molecule; B: Bolalipid molecule; C: Flexible string model of lipid chains in the hydrophobic part of bilayer membrane: effecitve string has incompressible area $A_0$, arbitrary conformation of fluctuating string is described  by function $\mathbf{R}(z)$ --- deviation from the straight line across the membrane thickness with coordinate $z$. 
} \label{1}
\end{figure}
\begin{figure}[tbp]
  \includegraphics[width=\linewidth]{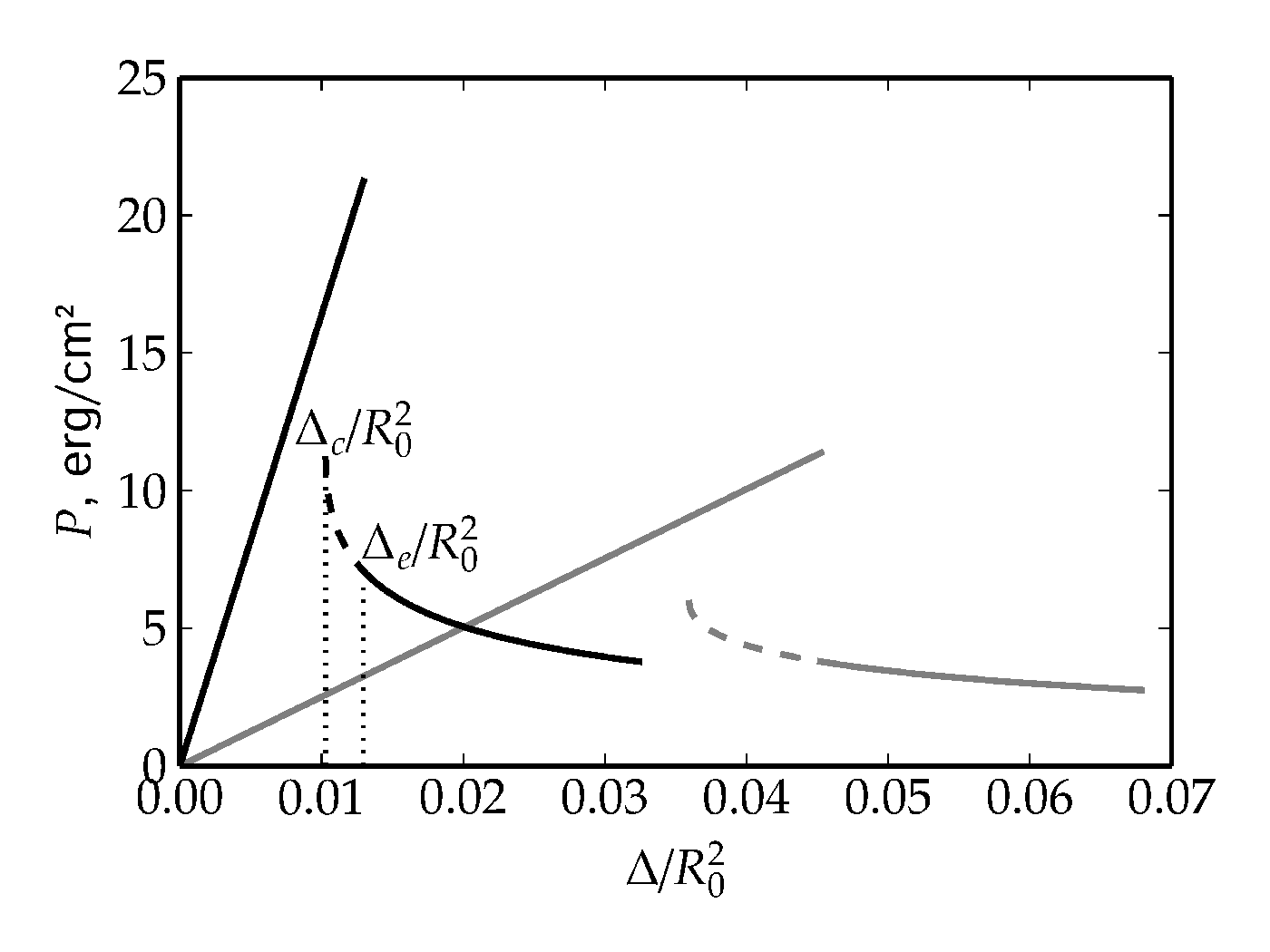}
  \caption{Fig. 2. Phase diagram of membrane with pore formation: P is lateral tension (pressure) in the membrane; $\Delta/R_0^2$ is area stretch of the membrane under external tension normalized with the initial undeformed membrane area. The dashed lines correspond to metastable states of the pore; higher pressure/slope curves - rigid rods; lower pressure/slope curves - flexible strings. Input parameters: membrane initial radius $R_0 = 95$ nm, $\gamma = 30$ erg/cm\textsuperscript{2}, $\gamma_R=20$ pN, $T = 300$ K; monopolar lipids: $N = 134200$ strings (in monolayer), $A_0 = 10~\AA^2$; bolalipids: $N = 80000$ rods, $A_0 = 27~\AA^2$.}
  \label{2}
\end{figure}

The purpose of the present work is calculation and comparison of the phase diagrams of porous fluid lipid membranes under lateral stretch deformation using different microscopic models developed recently \cite{mubao, mukhey}. We approach the problem of understanding the relatively high robustness of bolalipid membranes as compared with their monopolar lipid counterparts by considering distinction in their inter-chains entropic repulsion. Namely, the hydrocarbon chains of the monopolar lipid molecules are modeled by fluctuating semi-flexible strings (beams), and their bolalipid counterparts are modeled with straight rods, assuming the limit of higher bending (flexural) modulus. The models have been used to calculate the lateral pressure profiles and compressibility moduli of the monopolar lipid (see Fig.~\ref{1}, \textit{A}) and bolalipid membranes (see Fig.~\ref{2}, \textit{B}). The main outcome of the present derivation is analytical expressions for the critical lateral pressures that cause formation of finite radius pores  in  the monopolar lipid and bolalipid membranes. It is demonstrated that high chains flexural rigidity leads to a significant enhancement of the critical lateral area-stretching tension (lateral pressure), at which the membrane acquires a pore. Simultaneously, the radius of the critical pore is decreased by the stiffening of the chains (assuming pore edge energy stays approximately the same).  The Helmholtz free energy ($NVT$ - ensemble) is used in the derivations.

There is a phenomenological theory \cite{litster} of the pore formation under the constant tension and temperature. However in this theory a number of lipids doesn't hold, and the theory doesn't allow stable pore:
\begin{equation}
 \Delta F = 2 \pi R \cdot \gamma_R - \pi R^2 \cdot \tilde{\gamma}
  \label{eq:pt}
\end{equation}
Here $ \Delta F $ is the free enrgey change due to formation of a pore of radius $R$, $\gamma_R$ is the line tension of the pore edge, $\tilde{\gamma}$ is an effective surface tension of the membrane that includes the tension $\gamma$ at the membrane's hydrophilic-hydrophobic interface, as well as applied lateral tension that stretches the membrane. The first term describes energy of the pore edge, and the second term in Eq. (\ref{eq:pt}) describes gain in the elastic deformation energy due to pore formation. Once the radius $R$ of a pore becomes greater than $\gamma_R/\tilde{\gamma}$, the pore grows infinitely large. Hence, there are no stable pores, and one can't get critical pressure at which the smallest possible (meta)stable pore first appears.

Alternatively,  we consider the Helmholtz free energy change of the $NVT$ ensemble of lipids (constituting the membrane), being the sum of  the energy of the line tension of the pore edge and of the elastic deformation of the membrane at a given lateral area stretch \cite{farago_water-free_2003, tolpekina_simulations_2004}, which then developes  minima at finite values of the pore radius $R$: 
\begin{equation}
\Delta F = \frac{\pi K_a}{2} \frac{\left ( {R_\parallel}^{2} - {R_0}^{2} - R^2\right )^2}{{R_0}^{2}} + 2 \pi R \cdot \gamma_R
\label{eq:nvt}
\end{equation}
Here $K_a$ is the lateral stretch/compression modulus of the membrane, $R_\parallel$ is the radius of the outer circle that delimits the membrane with or without pore, $R_0$ is the radius of the membrane without pore under zero external lateral pressure, $\gamma_R$  is the line tension of the pore edge.

Below we derive Eq.~(\ref{eq:nvt}) from a microscopic flexible strings model \cite{mubao, mukhey}, and then calculate the phase diagram of the membrane considering bilayer of semi-flexible strings as a model of the lipid bilayer (formed by monopolar lipids), and monolayer of rigid rods as a model of the bolalipid membrane (see Fig.1). 
We write the Helmholtz free energy functional of a membrane as:
\begin{equation}
F = N \cdot F_t(A_p) + 2 \gamma \cdot \pi (R_\parallel^2 - R^2) + 2 \pi R \cdot \gamma_R \;.
\label{eq:helm-fsm}
\end{equation}
Here $N$ is number of hydrocarbon tails, $F_t(A_p)$ is the free energy of a tail, that sweeps an area $A_p$ in the membrane's plane. The  energy $F_t(A_p)$ allows for tail bending fluctuations and its collisions with the neighbors under the external lateral stretching stress. Also $\gamma$ is coefficient of the surface energy defined at hydrophilic-hydrophobic interface that usually separates lipid heads from hydrocarbon tails in the membrane. This surface tension is balanced by the entropic repulsion of the tails, resulting in zero overall tension of the self-assembled membrane:
\begin{equation}
  P_t - 2 \gamma = 0 \;.
  \label{eq:equ}
\end{equation}
Here and also in the Eq.~(\ref{eq:helm-fsm}) factor $2$ is due to the two surfaces of the membrane. 

The lateral increase of the membrane area $\pi (R_\parallel^2 - R_0^2 - R^2)$ equals:

\begin{equation}
  \pi (R_\parallel^2 - R_0^2 - R^2) = N \cdot \delta A \;.
  \label{eq:sur}
\end{equation}
where $\delta A$ is the increase of the area $A$ per single lipid.  Since lipid membrane is in the liquid crystalline phase the external lateral pressure is taken to be homogeneous across the membrane.
Substituting $\pi (R_\parallel^2 - R^2) = N \cdot \delta A +\pi R_0^2$ into (\ref{eq:helm-fsm}) and assuming change of area per lipid to be small: $A_p = A + \delta A$ one finds:
\begin{align}
  F =& N \left \lbrace F_t(A) + \delta A \cdot \frac{\partial F_t}{\partial A}(A) + \frac{\left (\delta A \right )^2}{2} \frac{\partial ^2F_t}{\partial A^2}(A) \right \rbrace + \nonumber \\
    & + 2 \gamma (N \cdot \delta A + \pi R_0^2) + 2 \pi R \cdot \gamma_R \;,
  \label{eq:F-series}
\end{align}
here $A$ is an area per lipid in a membrane with no pore under zero external pressure.
Recalling the equilibrium condition Eq.~(\ref{eq:equ}) and noting that 
\begin{equation}
P_t = - \dfrac{\partial F_t}{\partial A} 
  \label{eq:P_t}
\end{equation}
we find that:
\begin{equation}
  \frac{\partial F_t}{\partial A}(A) + 2 \gamma = 0 \;.
  \label{eq:A}
\end{equation}

Now introducing the energy of the membrane without pore under zero external lateral pressure:
\begin{equation}
  F_0 = N \cdot F_t(A) + 2 \gamma \cdot \pi R_0^2 \, ,
  \label{eq:E_0}
\end{equation}
and using Eqs.~(\ref{eq:A}), (\ref{eq:E_0}) and definition of $R_0$: $\pi R_0^2 = N \cdot A$, we can rewrite (\ref{eq:F-series}) as
\begin{eqnarray}
 \Delta F= &F - F_0 = N \cdot \dfrac{\left (\delta A \right )^2}{2} \dfrac{\partial ^2F_t}{\partial A^2} (A) + 2 \pi R \cdot \gamma _R \\
          & =\dfrac{\pi K_a}{2} \dfrac{\left ( {R_\parallel}^{2} - {R_0}^{2} - R^2\right )^2}{{R_0}^{2}} + 2 \pi R \cdot \gamma_R \;,
  \label{eq:F_s}
\end{eqnarray}
where
\begin{equation}
  K_{a} = A \cdot \frac{\partial^2 F_t}{\partial A^2}(A)\, ,
  \label{eq:K_a}
\end{equation}
is the area stretch/compressibility modulus calculated analytically in  \cite{mubao, mukhey}.
Hence, Eqs. (\ref{eq:F_s}) and (\ref{eq:K_a}) link free energy functional of the membrane with a pore with a microscopic theory for area compressibility modulus $K_a$ developed earlier  in \cite{mubao, mukhey}.

Next, following derivation of \cite{tolpekina_simulations_2004}, we extremize free energy given in Eq. (\ref{eq:F_s})  with respect to the pore radius $R$ : 
\begin{equation}
 {K_a} \left( R^3 - R\Delta \right) + \gamma_R {R_0^2}= 0
  \label{eq:der}
\end{equation}
where $\Delta = R_\parallel^2 - R_0^2$ is introduced for convenience. Then, the radius of  the pore, that minimizes the free energy of the membrane \cite{smirnov_course_1964,tolpekina_simulations_2004} is :
\begin{equation}
  R_m = 2 \sqrt{\dfrac{\Delta}{3}} \cos \frac{\phi}{3}
  \label{eq:R_m}
\end{equation}
where $\cos \phi = - \dfrac{\gamma_R R_0^2}{2 K_a} \left ( \dfrac{3}{\Delta} \right )^{3/2}$. The real valued solution $R_m$ exists, provided the discriminant of~(\ref{eq:der}) is negative:
\begin{equation}
  \frac{\gamma_R^2 R_0^4}{4 K_a^2} - \frac{\Delta^3}{27} < 0
  \label{eq:D<0}
\end{equation}
The area stretching tension $ P_{m}$ applied to a membrane with this pore is obtained by differentiation of  Eq. (\ref{eq:F_s}) under the condition~ (\ref{eq:der}): 
\begin{equation}
  P_{m} = -\frac{\partial (F - F_0)}{\pi \partial \Delta} = \frac{\gamma_R}{R_{m}}
  \label{eq:P_m}
\end{equation}

To create a pore in the pore-free membrane one applies a critical area stretching pressure $P_c$ by producing a critical area stretch  $\pi \Delta_c$, at which free energy $F - F_0$  in Eq.~(\ref{eq:F_s}) develops an inflection point  as a function of $R$ at $R = R_c$ ("critical pore's radius").  Hence,  $\Delta_c$ could be found by equating a discriminant of the Eq.~(\ref{eq:der}) to zero:
\begin{equation}
 \frac{\gamma_R^2 R_0^4}{4 K_a^2} - \frac{{\Delta_c}^3}{27}=0
  \label{eq:D=0}
\end{equation}
Using then Eqs. (\ref{eq:R_m}) and (\ref{eq:P_m}) we find consecutively all the critical parametrs \cite{tolpekina_simulations_2004}:
\begin{equation}
  \Delta_c = \frac{3}{2^{2/3}} \left ( \frac{\gamma_R R_0^2}{K_a} \right )^{2/3}, \; R_c = \left ( \frac{\gamma_R R_0^2}{2K_a} \right )^{1/3} \,,
  \label{eq:D_c}
\end{equation}
and
\begin{equation}
  P_c = \frac{\gamma_R}{R_c}= \left ( \frac{2\gamma_R^2 K_a}{R_0^2} \right )^{1/3} 
  \label{eq:P_c}
\end{equation}
But the critical pore is metastable. A smallest stable pore arises from it with the greater radius $R_e  =2^{2/3}R_c$, at which the free energy of the membrane  equals  the free energy of the membrane without pore: 
\begin{equation}
 \dfrac{\pi K_a}{2} \dfrac{\Delta^2}{{R_0}^{2}} = \dfrac{\pi K_a}{2} \dfrac{\left ( {\Delta} - R^2\right )^2}{{R_0}^{2}}+ 2 \pi R \cdot \gamma_R \;.
  \label{eq:stable}
\end{equation}
Solving  Eqs. (\ref{eq:stable}) and (\ref{eq:der}) simultaneously one finds parameters of the smallest possible stable pore:
\begin{equation}
 P_e = \dfrac{1}{2^{2/3}}P_c, \; R_e ={2^{2/3}}R_c,\; \Delta_e = {2^{1/3}}\Delta_c\,.
 \label{eq:D_e}
\end{equation}
Hence, Eq. (\ref{eq:D_e}) indicates that critical pressure exceeds the equilibrium pressure corresponding to the stable pore. Once a critical pore of the radius $R_c$ appears its radius increases up to $R_e$ until the area stretching pressure drops from the critical $P_c$ to the equilibrium value $P_e$. Simultaneously, the outer radius of the membrane increases monotonically. 
Now we link these phenomenological results with a microscopic models of lipid membrane by analytical derivation of the area stretching/compression modulus $K_a$. To follow this derivation we review some results of the flexible strings theory obtained previously \cite{mubao, mukhey}. 

In the semi-flexible strings model hydrophobic chain is treated as a string (beam) of a finite cross-section area $A_0$ (see Fig.1) and bending modulus $K_f$. Its deviations from a straight line along axis $z$ are considered as being small and the energy functional of the string is an integral over $z$:
\begin{equation}
  E_t = \int\limits_0^L \left [ \frac{\rho \dot{\mathbf{R}}^2 (z)}{2}  + \frac{K_f}{2} \left ( \frac{\mathrm{d}^2 \mathbf{R}}{\mathrm{d} z^2} \right )^2 + \frac{B \mathbf{R}^2}{2} \right ] {\, \mathrm{d} z}
  \label{eq:liq}
\end{equation}
The first term here is kinetic energy, the second term is bending energy and the last one models interaction between neighboring chains via entropic repulsion. 

The characteristic parameters of the string are: length, $L$ (for bolalipids we use $2 L$); number $N$ of $\mathrm{CH_2}$ groups of mass $m(\mathrm{CH_2})$.

The membrane is characterized by the hydrophobic surface tension $\gamma$ (see Eq.~(\ref{eq:equ})).

As the typical quantities for monopolar lipids we take $L = 15 \AA$, $A_0 = 10 \AA^2$, $N = 18$, $\gamma = 30$ erg/cm\textsuperscript{2} and $T = 300$ K for a temperature.

We estimate $K_f = k_B T L / 3$ \cite{mubao} using the Flory's formula for the bending rigidity of a polymer chain at room temperature. The chain density per unit of length  is estimated as: $\rho = m(\mathrm{CH_2}) N / L$. The self-consistent solution for parameter $B$ characterizing the entropic repulsion between chains  in ~(\ref{eq:liq}), at a particular average area $A$ swept by a lipid tail, is \cite{mubao, mukhey} :
\begin{equation}
  b = \frac{1}{4 \nu^{3/4} (\sqrt{a} - 1)^{8/3}} 
  \label{eq:bm}
\end{equation}
where dimensionless variables are introduced:
\begin{equation}
  a = \frac{A}{A_0}, ~~ b = B \cdot \frac{L^4}{K_f}, ~~ \nu = \frac{K_f A_0}{\pi k_B T L^3} \;,
  \label{eq:abn}
\end{equation}
and the limit $b\gg1$ is assumed.
The lateral pressure of a monolayer of the hydrocarbon chains is found by a substitution of ~(\ref{eq:bm}) into ~(\ref{eq:liq}) and differentiation of the free energy $F_t$ according to ~(\ref{eq:P_t}) \cite{mubao}  :
\begin{equation}
  \label{eq:P}
  P_t = \frac{k_B T}{3A_0 \nu^{1/3} \sqrt{a} (\sqrt{a} - 1)^{5/3}} \;.
\end{equation}
Using the balance equation for the monolayer: $P - \gamma = 0$, we find \cite{mubao, mukhey}:
\begin{equation}
  \frac{1}{\sqrt{a} (\sqrt{a} - 1)^{5/3}} = {3 \nu^{1/3}} \frac{A_0 \gamma}{k_B T}\equiv g.
  \label{eq:g}
\end{equation}
The typical lipid chain parameters lead to an estimate: $\nu\approx 0.01$ and $g\approx 0.4$.  Since we consider monopolar lipid membrane in a liquid state, i.e. $A\gg A_0$ or equivalently $a\gg 1$, we choose $g \ll 1$ limit to obtain analytical expressions, and thus find from above:
\begin{equation}
  a = g^{-3/4}\,.
  \label{eq:am1}
\end{equation}
Now we derive $K_a$ defined in ~(\ref{eq:K_a}), using ~(\ref{eq:P_t}), (\ref{eq:P}) and ~(\ref{eq:g}), (\ref{eq:am1}):
\begin{equation}
 K_{a\gg1}=-2A \frac{{\partial}P_t}{{\partial} A} = \frac{8 k_B T}{9 A_0 \nu^{1/3} (\sqrt{a} - 1)^{8/3}}\approx   \frac{8\gamma}{3}
  \label{eq:d2Fa_gg1}
  \end{equation}
  
To study the opposite limit, $a\rightarrow 1$, we consider an ultimate case of tightly packed hydrocarbon chains ($a\rightarrow 1+0$) using model of rigid rods, $K_f = \infty$. Then, curved conformations of the chains have infinite energy and therefore drop out from the energy functional:
\begin{equation}
  E_t =  \int\limits_0^{2 L} \left [ \frac{\rho \dot{\mathbf{R}}^2 (z)}{2} + \frac{B \mathbf{R}^2}{2} \right ] {\, \mathrm{d} z}\,.
  \label{eq:E_b}
\end{equation}
Here $2L$ is thickness of the hydrocarbon part of the membrane. Despite the rod is rigid, a deviation $\mathbf{R}(z)$ might not be zero, since it also includes movements of the rod as a whole in the lateral directions. Then, operator of the potential energy, $\hat H \equiv B$, has only single (constant) eigenfunction, which we normalize:
\begin{equation}
  \int_0^{2L} R_0^2 (z) {\, \mathrm{d} z} = 1 ~~ \Rightarrow ~~ R_0(z) = \frac{1}{\sqrt{2L}}
  \label{eq:R_0}
\end{equation}
All the formalism of the flexible strings model holds \cite{mubao, mukhey} and one finds: 
\begin{equation}
  B(a) = \frac{k_B T \pi}{2 L A_0 (\sqrt{a} - 1)^2}\,,
  \label{eq:Bb}
\end{equation}
that gives:
\begin{equation}
  P_t = \frac{k_B T}{A_0} \frac{1}{\sqrt{a} (\sqrt{a} - 1)}
  \label{eq:Ptb}
\end{equation}
Using balance equation ~(\ref{eq:equ})  we obtain instead of ~(\ref{eq:am1}):
 \begin{equation}
  a = 1 + \frac{\epsilon}{2};\; \epsilon\equiv \frac{k_B T}{A_0 \gamma}\ll 1\,,
  \label{eq:Ba0}
\end{equation}
where the limit $\epsilon \ll 1$ has to be assumed for $a\rightarrow 1$ to be true \footnote{This inequality could arise e.g. due to morphology of the particular bolalipid molecules built with ether linkages \cite{rosa_structure_1986} unlike monopolar lipids, that are built with ester linkages.}. 
We then calculate ${{\partial}^2 F_t}/{\mathrm{\partial} A^2}$ in order to obtain $K_a$  from ~(\ref{eq:K_a}) :
\begin{equation}
  K_{a\approx 1}=-a\frac{{\partial}P_t}{{\partial}a} = 8\gamma \frac{A_0 \gamma}{k_B T}\equiv\frac{8\gamma}{\epsilon} \;(\gg \gamma).
  \label{eq:d2F-bola}
\end{equation}
This result is remarkable, since direct comparison with Eq. (\ref{eq:d2Fa_gg1}) indicates that in the limit $a\rightarrow 1$ , due to $\epsilon\rightarrow 0$, the lateral compressibility coefficient $K_a$ is enhanced by $1/\epsilon\gg1$ times with respect to the case $a\gg 1$ at $\epsilon\geq 1$ . 

Finally, we substitute the microscopic theory results ~(\ref{eq:d2Fa_gg1}), (\ref{eq:d2F-bola}) into the general relations ~(\ref{eq:D_c}), (\ref{eq:P_c}) and (\ref{eq:D_e}) and find parameters characterizing critical (and equilibrium pores, see relations in Eq. (\ref{eq:D_e})) in "liquid disordered" ($a\gg 1$) and "liquid ordered" ($a\rightarrow 1$)\footnote{These names are of course rather tentative} membranes:  
\begin{eqnarray}
 P_c^{a\gg 1}&=&\left( \frac{16\gamma\gamma_R^{2}}{3R_0^2}\right)^{1/3}= \left(\dfrac{\epsilon}{3} \right)^{1/3}P_c^{a\approx 1} ;
     \\      R_c ^{a\gg1}&=&\left( \frac{3\gamma_RR_0^2}{16\gamma}\right)^{1/3}=  \left(\dfrac{\epsilon}{3} \right)^{-1/3}R_c^{a\approx 1};
 \label{eq:be}
\end{eqnarray}
and $\Delta_c = 3R_c^2$, and $\epsilon\ll 1$ is defined in ~(\ref{eq:Ba0}).
These results indicate, that stiffening  of the lipid chains would lead to enhancement of the critical stretching tension (pressure) for pore formation, but simultaneously, would decrease the radius of the thus formed pore. Hence, the membrane becomes more robust to external mechanical lateral stress. The phase diagram of the membranes in the "liquid disordered" and "liquid ordered" limits (i.e. in $a\ll1$ and $a\rightarrow 1$ states, correspondingly) is plotted in Fig. 2 in the form $P=P_m(\Delta)$ using Eqs. (\ref{eq:D<0}) and (\ref{eq:P_m}).

To summarize, we had calculated critical lateral tensions and corresponding pore radii for monopolar lipid and bolalipid membranes. Our results for two distinct microscopic models of hydrophobic lipid chains: semi-flexible strings and rigid rods suggest that stiffening the chains leads to enhancement of membrane lateral robustness. 

We acknowledge partial support by the RFFI-KOMFI grant Nr.\ 13\ch 04\ch 40327\ch N and NUST MISIS infrastructure grant.

\end{document}